\documentclass[preprint]{elsarticle}

\usepackage{lineno,hyperref}
\usepackage{amssymb}
\usepackage{amsmath}
\usepackage{multirow}

\journal{Physics Letters A}

\begin{document}

\begin{frontmatter}

\title{Pushing the limits of EPD zeros method}

\author[1]{R. G. M. Rodrigues\texorpdfstring{\corref{cor1}}{}}
\address[1]{Laborat{\'o}rio de Simula{\c c}{\~a}o, Departamento de F{\'i}sica, ICEx, Universidade Federal de Minas Gerais, 31720-901 Belo Horizonte, Minas Gerais, Brazil.}
\cortext[cor1]{Corresponding author.}
\ead{ronaldogmr@ufmg.br}

\author[1]{B. V. Costa}
 \ead{bvc@fisica.ufmg.br}

\author[1]{L. A. S. M{\'o}l}
 \ead{lucasmol@fisica.ufmg.br}

\begin{abstract}
The use of partition function zeros in the study of phase transition is growing in the last decade mainly due to improved numerical methods as well as novel formulations and analysis. In this paper the impact of different parameters choice for the energy probability distribution (EPD) zeros recently introduced by Costa et al is explored in search for optimal values. Our results indicate that the EPD method is very robust against parameter variations and only small deviations on estimated critical temperatures are found even for large variation of parameters, allowing to obtain accurate results with low computational cost. A proposal to circumvent potential convergence issues of the original algorithm is presented and validated for the case where it occurs.
\end{abstract}

\begin{keyword}
Phase Transition, Potts Model, Ising Model, Function Partition Zeros, EPD Zeros, Fisher Zeros.
\end{keyword}

\end{frontmatter}


\section{Introduction}

Phase transition is one of the core subjects in statistical physics. Several interesting behaviors such as superconductivity, ferromagnetism, and many others, emerge in systems after a phase transition. The concept of phase itself is somewhat difficult to broadly define but one can easily identify it in the presence of phase boundaries or phase coexistence. At those boundaries, where a phase transition occurs, it is well known that the system's free energy has a non-analytic behavior which leads to divergences or discontinuities in its derivatives, i.e., in a phase transition it is expected to see quantities like magnetization and specific heat presenting discontinuities or divergences. To study these behaviors, several approximated methods were developed, Mean field approximations \cite{kadanoff2009more}, the renormalization group \cite{RevModPhys.70.653}, Monte Carlo \cite{landaubinder2014} method and finite size scaling theory \cite{finite.size} are some of the most used approaches. In addition to those, Yang-Lee \cite{PhysRev.87.404} introduced a different way to study phase transitions. They rewrote the grand canonical partition function as a polynomial in the complex fugacity plane and showed that the polynomial's roots have all the systems' thermodynamic information, such as the partition function itself. Furthermore, they showed that at the thermodynamic limit the density of roots completely determines the system critical behavior accumulating near phase transition points; those roots are called Yang-Lee zeros. The theory of zeros developed by Yang-Lee is among the few rigorous theories regarding phase transitions and is considered a cornerstone of statistical physics.
    
	 The concept of Yang-Lee zeros was extended by Fisher \cite{Fisher1} to the canonical ensemble, where the canonical partition function was rewritten as a polynomial in the complex temperature plane. However, solving this polynomial showed to be a challenge. Even for simple systems the polynomial has a high degree easily surpassing four or more orders of magnitude. Its coefficients are given by the density of states, a quantity difficult to find and known exactly only for a few systems. Moreover, for most systems, the density of states has huge values making the polynomial even more difficult to deal with. Because of those characteristics, even for the state of the art root finder algorithms, one may expect several numerical instabilities.

  The caveat of the Fisher zeros polynomial raised above was solved by Costa, Mól and Rocha in \cite{Costa2016a}, where the polynomial was simplified after a transformation. The main advantage of this method lies in the fact that the transformed polynomial has as coefficients the energy probability density (EPD), being possible to judiciously discard coefficients regarding states with low probability to occur without losing relevant thermodynamic information, drastically reducing the polynomial degree. 

\begin{figure}
 \begin{center}
    \includegraphics[width=0.49\columnwidth]{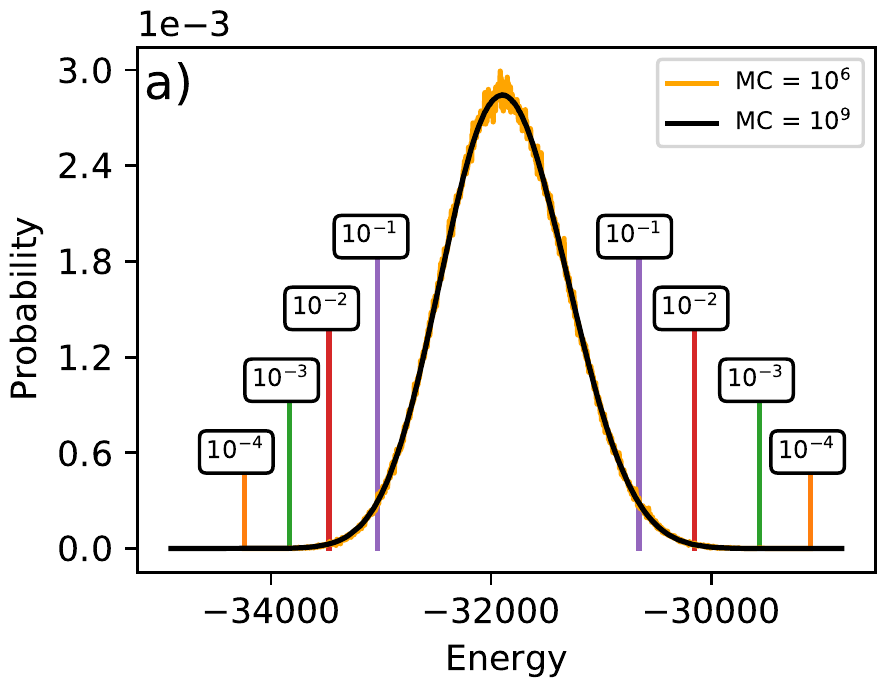} \\
 \end{center}
 \includegraphics[width=0.49\columnwidth]{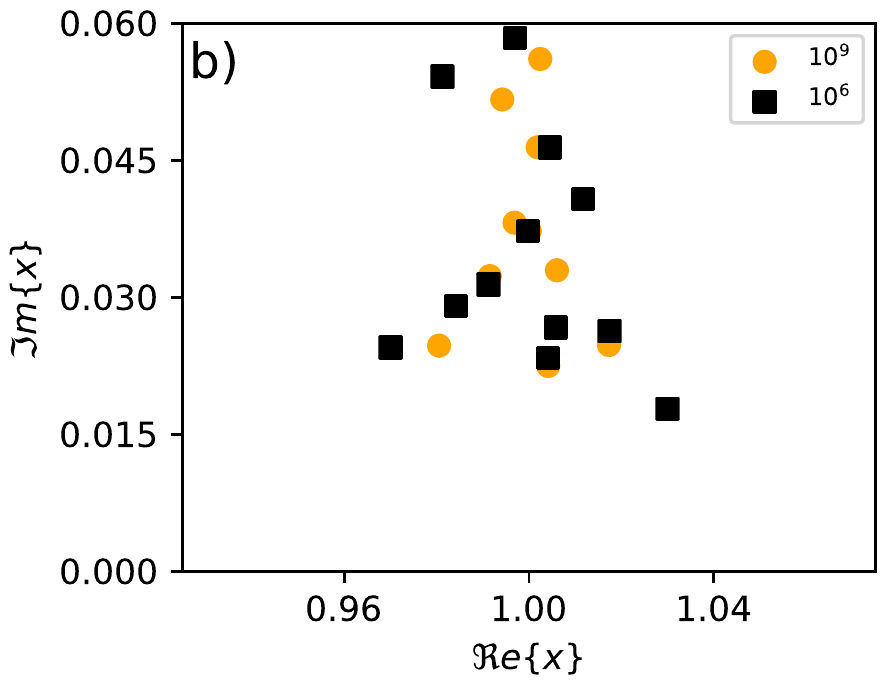}
 \includegraphics[width=0.49\columnwidth]{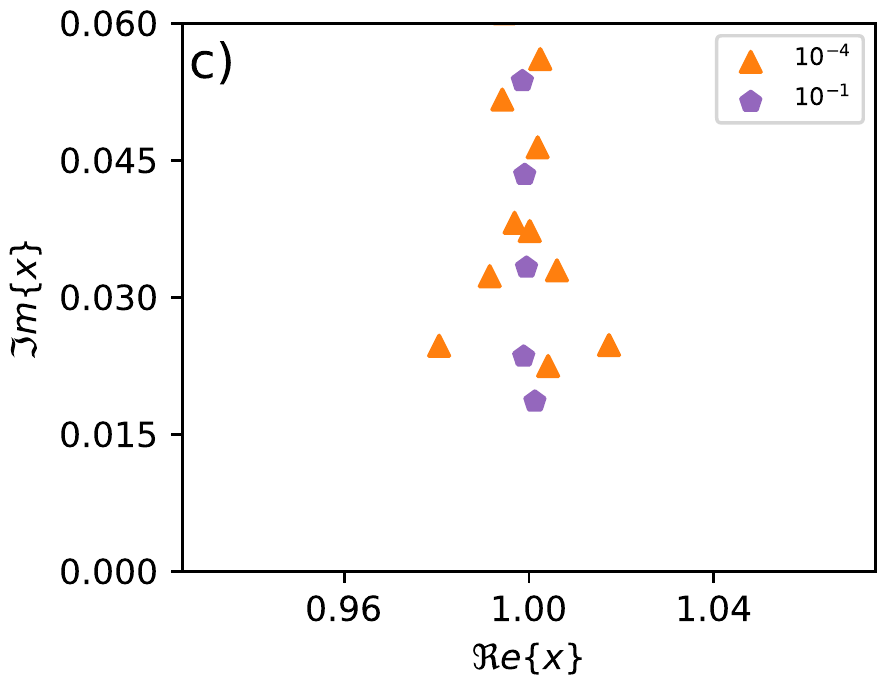}
 \caption{a) Two EPD for a square lattice Ising model with lattice size $L=150$ at temperature $T = 2.2700 J/k_b$. The vertical lines indicate four different discard thresholds $10^{-1}, 10^{-2}, 10^{-3}, 10^{-4}$. For each threshold only states between the marked vertical lines are considered as coefficients of the polynomial. b) Map of zeros made using the EPD for $10^9$ and $10^6$ MCS without discarding any coefficient. c) Map of zeros made using the EPD with $10^9$ MCS and with discard thresholds of $10^{-4}$ and $10^{-1}$. Although the map of zeros in b) and c) are different, they represent the same system, and the EPD method should give results statistically equivalent to each other.}
 \label{problem}
\end{figure}

  Although the EPD zeros were successfully applied to the study of phase transitions, such as discontinuous, continuous and topological ones \cite{Costa2016a, Lima2019, Costa2019}, some points still need to be clarified. Since the EPD method is highly dependent on the zero's positions, changes in the polynomial's coefficients such as coefficient discards or statistical fluctuations when estimating the EPD, may have a significant impact in quantities found by the method such as the critical temperature or critical exponent. To exemplify the problem, figure \ref{problem} shows EPDs and their respective map of zeros for different coefficient discard thresholds and different Monte Carlo steps (MCS) used to build the EPD. Although the map of zeros shown is different for each choice of coefficients discard threshold and number of Monte Carlos Steps, they all represent the same system and should give results for critical temperature and critical exponents statistically equivalents. With this in mind, in this contribution we address  the effects of choosing different coefficient discard thresholds and MCS when using the EPD zeros approach in systems displaying continuous and discontinuous phase transition. In what follows we briefly describe the Fisher and EPD zeros methods and the models we considered. Results for the Ising model and the six-state Potts model are shown in sections \ref{sec:res-ising} and \ref{sec:potts}, respectively. A convergence issue found for the Ising model is discussed in section  \ref{sec:prob} and its solution in section \ref{sec:solv}. The conclusions and final remarks are drawn in section \ref{sec:conc}.

\section{Fisher zeros}

	Following the ideas of Yang-Lee \cite{PhysRev.87.404}, Fisher \cite{Fisher1} considered an analytic continuation of the temperature to the complex plane and rewrote the canonical partition function $Z$ as a polynomial,

\begin{equation}\label{z}
Z = \sum_E g(E)e^{-\beta E} = e^{-\beta \varepsilon_o} \sum_n g_nz^{n} \hspace{0.5cm} \text{,}
\end{equation}
where $z=e^{-\beta \varepsilon}$, $\beta = 1/k_B T$ is the inverse temperature, $k_B$ is the Boltzmann constant, $g(E)$ is the density of states, the energy $E$ is discretized by setting it to $E_n = \varepsilon_o + n \varepsilon$ with $n \in \mathbb{N}$, $\varepsilon_0$ and $\varepsilon$ being constants and $g_n=g(E_n)$ \cite{ROCHA201688}. Since the polynomial's coefficients are positive, $g(E) \geq 0$, all roots appear as conjugate pairs, and none of them are in the positive real axis. Only in the thermodynamic limit, at points where a phase transition occurs, it is expected to see a real positive root $z_c$; these roots are called dominant zeros or leading zeros.

\section{\label{sec:epd}EPD zeros}

	The EPD zeros are obtained by multiplying Eq. \ref{z} by $e^{-\beta_o E}e^{\beta_o E} = 1$. Then,
	
\begin{equation}
Z = \sum_E g(E)e^{-\beta_o E}e^{-(\beta-\beta_o)E} = e^{-\Delta \beta \varepsilon_o} \sum_n h_n(\beta_o) x^n \text{,}
\end{equation}
where $\beta_o = 1/k_b T_o$ is an arbitrary inverse temperature, $\Delta \beta = \beta - \beta_o$, $h(E,\beta_o) = g(E)e^{-\beta_o E}$ is the non-normalized energy probability distribution function at $\beta_o$ and $x=e^{-\Delta \beta \varepsilon}$. We notice this corresponds to a rescaling of the Fisher zeros since $x = e^{-\beta_o \varepsilon} z$. The main improvement is that since states with low probability to occur have little relevance to the system thermodynamics at $\beta_o$ they can be discarded. By doing that, the polynomial's degree is drastically reduced. Furthermore, as for the Fisher zeros in the thermodynamic limit, the dominant EPD zero approaches the real axis, touching it signalizing that a phase transition occurred.

  The EPD zeros have another important characteristic, in the thermodynamic limit, if we choose $\beta_o = \beta_c$, where $\beta_c$ is the critical inverse temperature, the dominant zero $x_c$ will be always at the point $(1,0)$. Moreover, for a $\beta_o$ not so close to $\beta_c$, the distance from the dominant zero to the point $(1, 0)$ can be used to estimate how far $\beta_o$ is from $\beta_c$. Therefore, the following algorithm was proposed to iteratively approximate $\beta_o$ to $\beta_c$ even for finite systems.

\begin{itemize}
    \item[1] Build a single EPD $h_n(\beta_o^j)$ at $\beta^j_o$.
    \item[2] Normalize $h_n(\beta_o^j)$, so its maximum value is $1$, and discard coefficients smaller than a given threshold.
    \item[3] Find the zeros of the polynomial with coefficients given by $h_n(\beta_o^j)$.
    \item[4] Find the dominant zero, $x^j_c$.
    \begin{itemize}
        \item[a)] If $x^j_c$ is close enough to the point $(1, 0)$, stop.
        \item[b)] Else, make $\beta^{j+1}_o = - \frac{\text{ln}(\Re[x^j_c])}{\varepsilon} + \beta^j_o$ and go back to $(1.)$.
    \end{itemize}
\end{itemize}
    
	The main idea behind it is that the point $(1,0)$ may behave as an attraction point for systems with a phase transition when coefficients are discarded and finite systems are considered. In this way, as close to the transition point the EPD is drawn, more precise the estimate of the transition temperature is, since for reasonable choices of coefficients discard thresholds only negligible information for the phase transition are disregarded.

\section{Models}

  In this work, we consider two very well known spin models that serve as benchmarks in the study of continuous and discontinuous phase transitions, the Ising and Potts models, that we briefly present in what follows.
  
  The Ising model Hamiltonian is given by,
  
\begin{equation}
H = -J\sum_{<i,j>} \sigma_i\sigma_j \text{,}
\end{equation}
where $J>0$ is a coupling constant, $\sigma_i$ is a spin at site $i$ with value $\pm 1$ and $<i,j>$ means a sum over nearest neighbors. The 2D version of this model in a square lattice has a continuous phase transition at $T_c = 2 / \text{ln}(1+\sqrt{2}) J/k_b$ and a critical exponent $\nu = 1$ \cite{potts.review}.

  The q-states Potts model Hamiltonian is given by,
  
\begin{equation}
H = -J\sum_{<i,j>}\delta(\sigma_i,\sigma_j) \text{,}
\end{equation}
where $\sigma_i$ is a spin on the $i\text{-}th$ site that assumes discrete values in range $[1,...,q]$, $J>0$ is a coupling constant, $\delta$ is the Kronecker delta, and the sum is taken over the nearest neighbors. This model presents a continuous and a discontinuous phase transition for $q \leq 4$ and $q>4$ respectively in two dimensions. Its critical temperature is known exactly to be $T_c = 1 / \text{ln}(1+\sqrt{q}) J/k_b$ \cite{potts.review}. In special, for $q=2$ this model can be mapped into the Ising model by a simple energy rescaling.

\section{Results}

	The simulations were carried out as follows. For the Ising model, a single spin flip process was used and three temperatures close to the critical one were simulated using the Metropolis algorithm for lattices with lateral size ranging from $L=90$ to $L=180$. For the Potts model, the Wolff \cite{wolff} algorithm was used in addition to the single spin flip and fifteen temperatures very close to the critical one were simulated, for lattices with lateral size ranging from $L=90$ to $L=180$. Moreover, multiple histograms reweighting technique \cite{PhysRevLett.63.1195} was applied to create the EPDs used in the algorithm described in section \ref{sec:epd}.
	
	As will be described in more detail in the next section, to estimate the critical temperature and critical exponent, it is necessary to find the mean value of $T_c(L)$ and $\Im m(x_c(L))$ for each lattice size $L$. Therefore, for each lattice size, 5 different simulations were made at each temperature. Then, the multiple histograms reweighting technique together with the algorithm described in section \ref{sec:epd} was used to find the values of $T_c(L)$ and $\Im m(x_c(L))$. With these values, it is straightforward to find the mean and the errors of $T_c$ and $\Im m(x_c)$ for each lattice size. This process was used for the Ising model and the six-state Potts model using $10^{6}$, $10^{7}$ and $10^{8}$ MCS with discard threshold ranging from $10^{-100}$ to $10^{-1}$.

    In sections \ref{sec:res-ising} and \ref{sec:potts}, we present tables summarizing the results obtained for $T_c$ and $\nu$ with different discard thresholds and different MCS. Furthermore, images shown in the next sections, unless otherwise stated, were obtained using $10^6$ MCS for each simulated temperature and with lattice size of $L=150$.
     
\subsection{Ising model}

\subsubsection{\label{sec:res-ising}Critical temperature and critical exponent}

\begin{figure}[ht]
 \includegraphics[width=0.49\columnwidth]{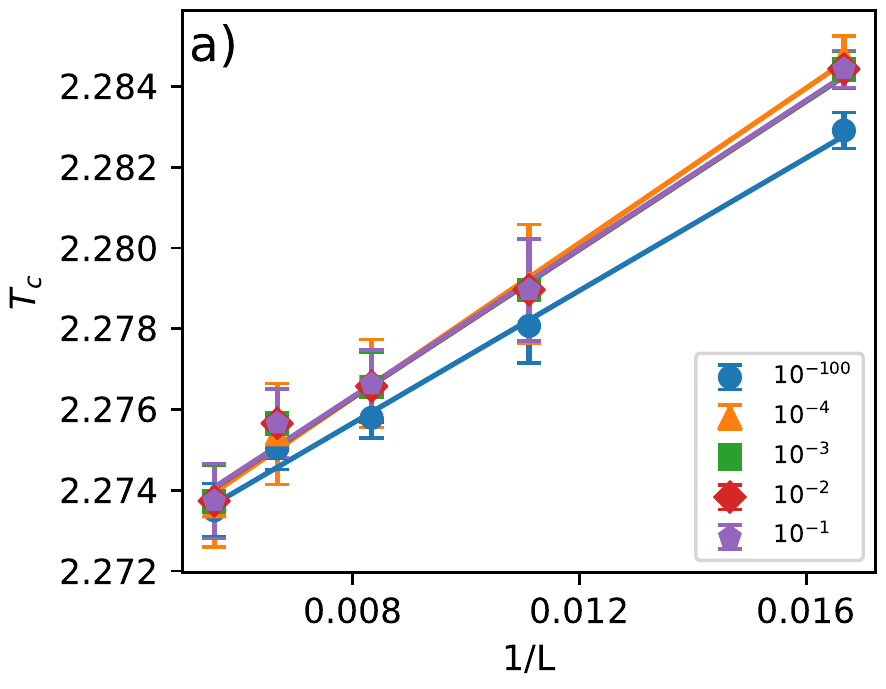}
 \includegraphics[width=0.49\columnwidth]{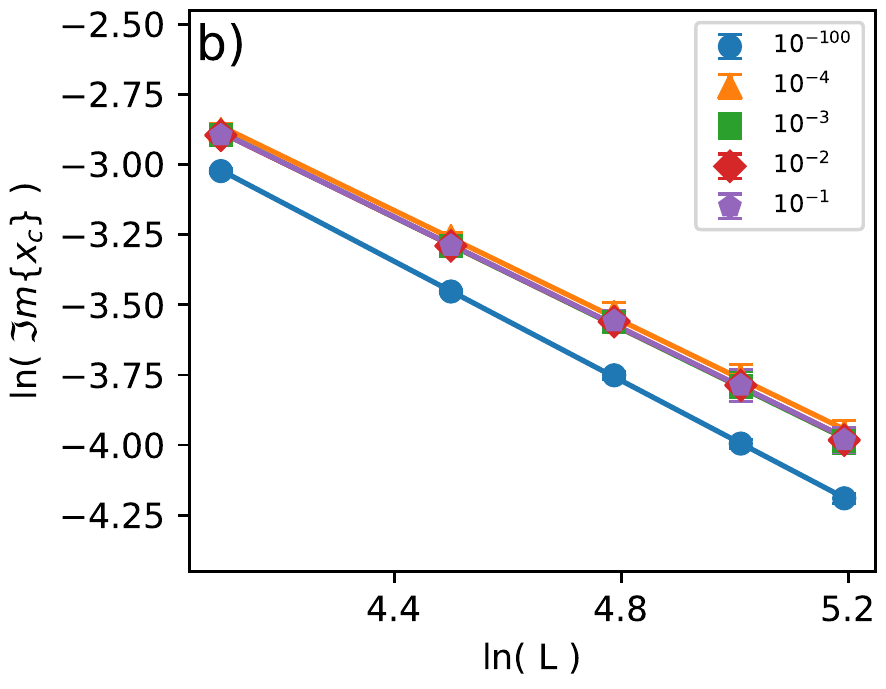}
 \caption{Typical regression to find a) $T_c$ and b) $\nu$ for thresholds ranging from $10^{-1}$ to $10^{-100}$. The data obtained was made using $10^{6}$ MCS and lattices of size $L=90,120,150,180$. All the values found for $T_c$ and $\nu$ were shown in tables \ref{tab:Is-tc} and \ref{tab:Is-nu} }
 \label{ising-regression}
\end{figure}

Estimates of the critical temperature were found using the following finite size scaling equation:
\begin{equation}\label{reg_t}
    T_c(L) \sim T_c + aL^{-1/\nu},
\end{equation}
and since the imaginary part of the dominant zero goes to zero in the thermodynamic limit, the following scaling equation is used to estimate the critical exponent, 
\begin{equation}\label{reg_im}
    \Im m (x_c(L)) \sim bL^{-1/\nu}.
\end{equation}
	The data used in these estimates were obtained by an automated application of the original algorithm, corrected for a convergence issue we found and discuss in the following sections.
	The results obtained for $T_c$ and $\nu$ were summarized in tables \ref{tab:Is-tc} and \ref{tab:Is-nu} for several combinations of thresholds and MCS. Moreover, a typical regression for the critical temperature and critical exponent is shown in figure \ref{ising-regression}, for the sake of clarity, the figure shows only the regression with $10^6$ MCS. As can be seen in tables \ref{tab:Is-tc} and \ref{tab:Is-nu}, all the results are close to the expected ones, $\nu=1$ and $T_c=2.2691 J/k_b$, independent of the threshold and the MCS chosen.

\begin{table}[ht]
\centering
\begin{tabular}{l|ccc}
 \multicolumn{1}{c|}{Threshold}&\multicolumn{3}{c}{MCS} \\
  & $10^{6}$ & $10^{7}$ & $10^{8}$\\ \hline
 $10^{-100}$ & $2.2689(4)$ & $2.2697(3)$ & $2.2696(2)$ \\
 $10^{-4}$ & $2.2689(4)$ & $2.2697(3)$ & $2.2696(2)$\\
 $10^{-3}$ & $2.2689(4)$ & $2.2697(3)$ & $2.2696(2)$ \\
 $10^{-2}$ & $2.2684(3)$ & $2.2696(3)$ & $2.2695(2)$ \\
 $10^{-1}$ & $2.2691(4)$ & $2.2697(2)$ & $2.2696(2)$ \\
\end{tabular}
\caption{Results obtained for the critical temperature of the 2D Ising model using different combinations of threshold and Monte Carlo steps.}
\label{tab:Is-tc}
\end{table}

\begin{table}[ht]
\centering
\begin{tabular}{l|ccc}
 \multicolumn{1}{c|}{Threshold}&\multicolumn{3}{c}{MCS} \\
  & $10^{6}$ & $10^{7}$ & $10^{8}$\\ \hline
 $10^{-100}$ & $0.98(1)$   & $0.993(5)$  & $0.9987(9)$ \\
 $10^{-4}$ & $0.98(1)$   & $0.994(5)$  & $1.000(1)$\\
 $10^{-3}$ & $0.99(1)$   & $0.998(4)$  & $1.0027(7)$ \\
 $10^{-2}$ & $0.98(1)$   & $0.98(2)$   & $0.986(4)$ \\
 $10^{-1}$ & $1.0613(4)$ & $1.0635(2)$ & $1.0666(6)$ \\
\end{tabular}
\caption{Results obtained for the critical exponent $\nu$ of the 2D Ising model using different combinations of threshold and Monte Carlo steps.}
\label{tab:Is-nu}
\end{table}

\subsubsection{\label{sec:prob}Convergence issue and the dominant zero}

\begin{figure}[ht]
 \includegraphics[width=0.49\columnwidth]{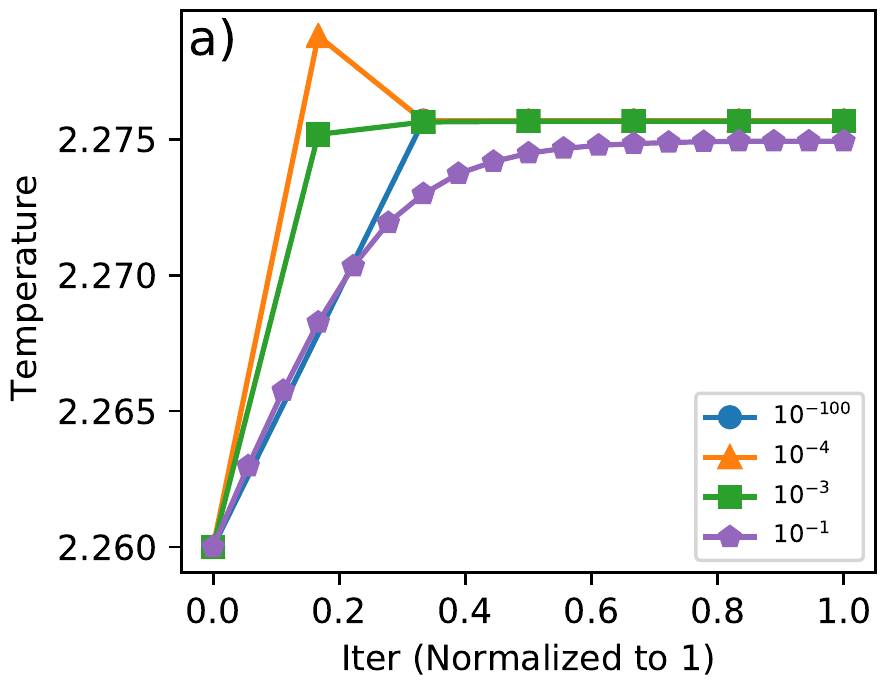}
 \includegraphics[width=0.49\columnwidth]{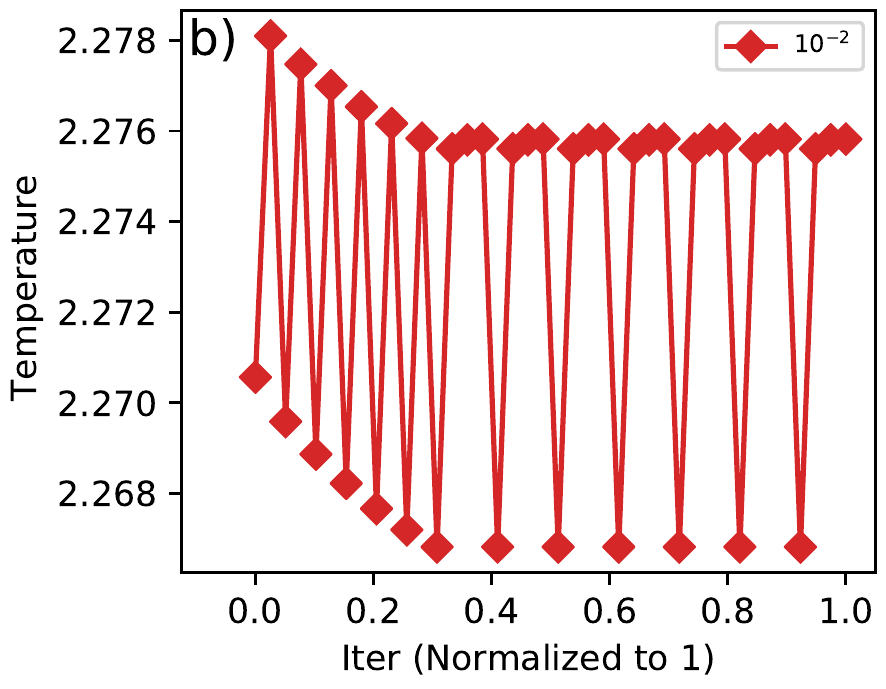}
 \caption{a) A typical process of convergence for the discard threshold of $10^{-1}, 10^{-3}, 10^{-4}$ and $10^{-100}$. b) Oscillation pattern characteristic of the algorithm incorrectly selecting the dominant zero found for the discard threshold of $10^{-2}$.}
 \label{conv-and-non-conv}
\end{figure}

    Considering all the partition function zeros, the dominant zero is the most important one for identifying a phase transition. However, find it among other zeros may not be an easy task for automated applications of the algorithm. Indeed, automated applications rely simply on the identification of the dominant zero as the one closer to the point (1,0) regardless of other factors. Here we found that there are some pathological situations where zeros that are not determinant to the phase transition may become closer to the point $(1,0)$ than the dominant zero. As a consequence, the algorithm convergence to the transition temperature is affected, since non-dominant zeros may be confused with the dominant one. As an example, in figure \ref{conv-and-non-conv}a it is shown the algorithm convergence for several coefficient discard thresholds and in figure \ref{conv-and-non-conv}b an oscillating pattern caused by the algorithm incorrectly selecting the dominant zero. It is important to highlight that systems with lattice of size $L=[90,120,150,180]$ were used, but only for $L\geq 150$ the convergence issue was observed.

\subsubsection{\label{sec:solv}Solving the convergence issue}

\begin{figure}
 \includegraphics[width=0.49\columnwidth]{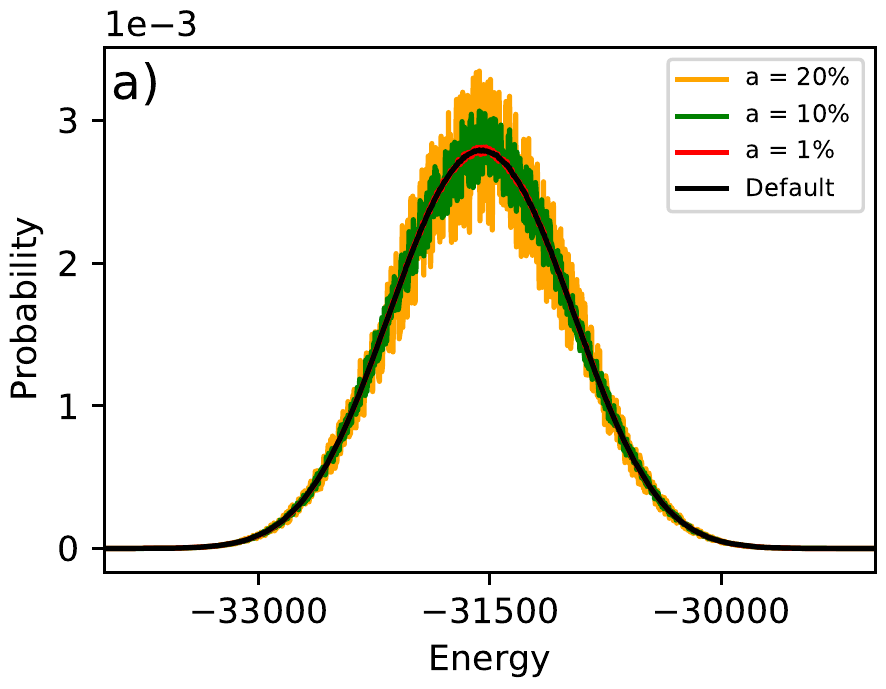}
 \includegraphics[width=0.49\columnwidth]{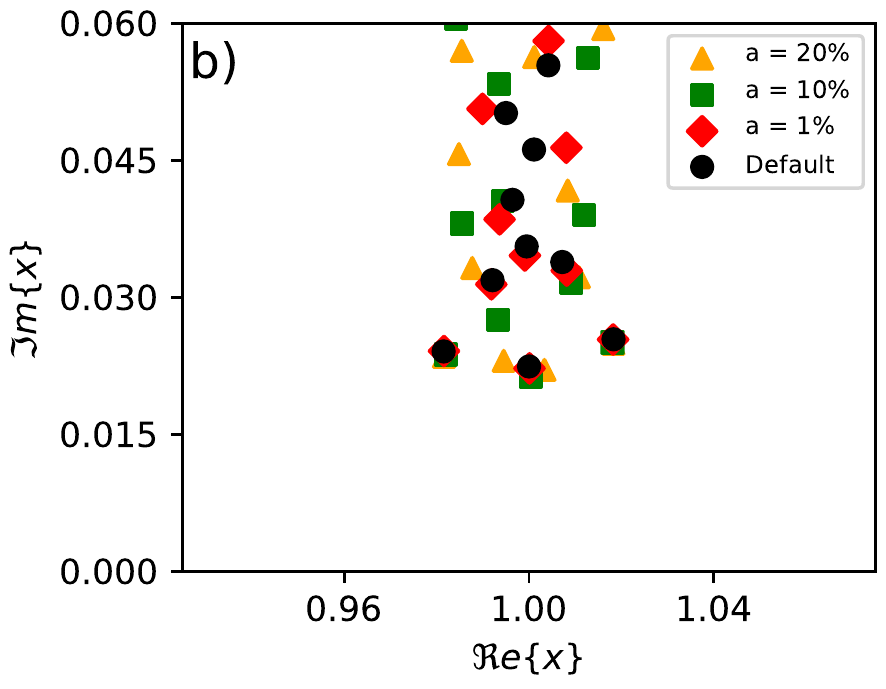}
 \caption{$a)$ An EPD made with $10^8$ MCS (black) where it was applied a random perturbation on it following the equation \ref{pert} with $a=[1\%,10\%,20\%]$. Although the perturbation changes all coefficients values, the dominant zero is stable $b)$ while other zeros change their position. To find the zeros in $b)$ a threshold of $10^{-4}$ was used.}
 \label{est}
\end{figure}

\begin{figure}
 \includegraphics[width=0.49\columnwidth]{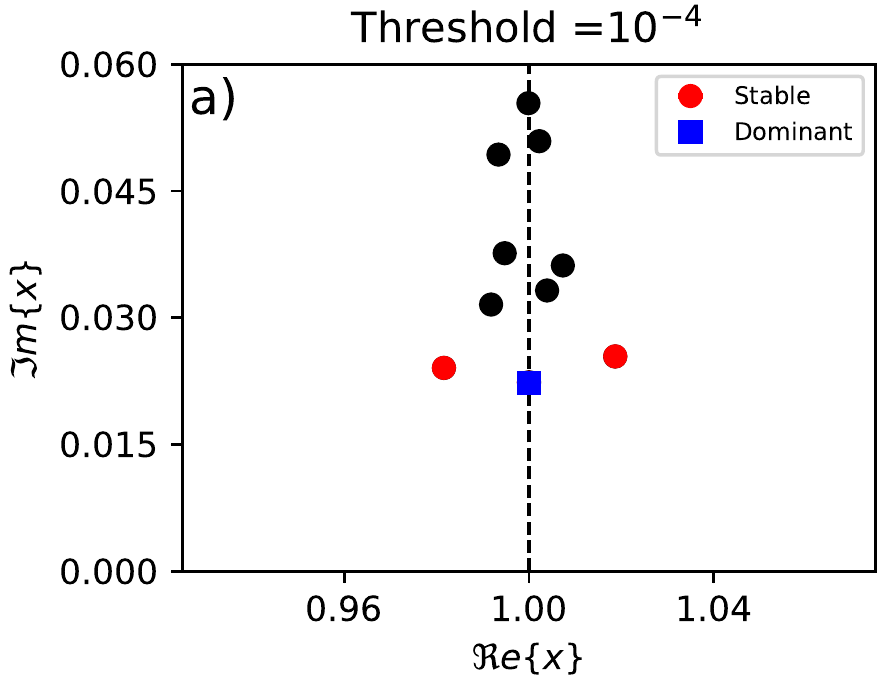}
 \includegraphics[width=0.49\columnwidth]{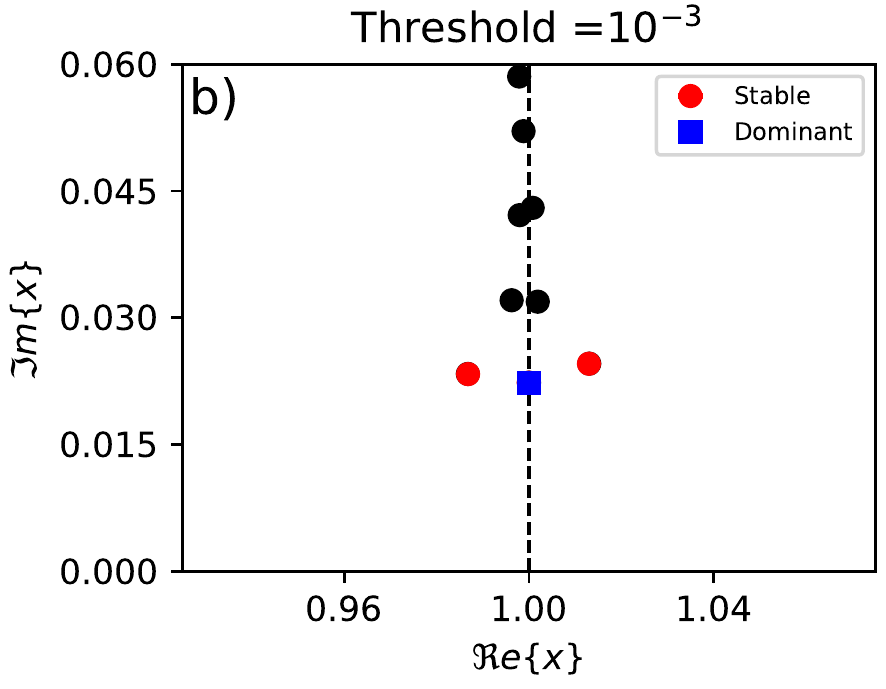}
 \includegraphics[width=0.49\columnwidth]{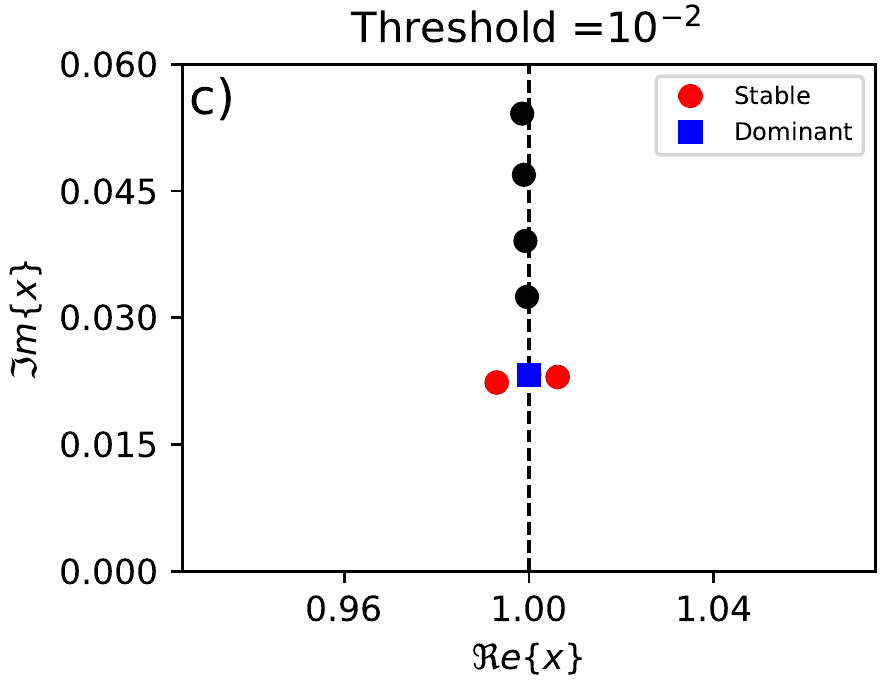}
 \includegraphics[width=0.49\columnwidth]{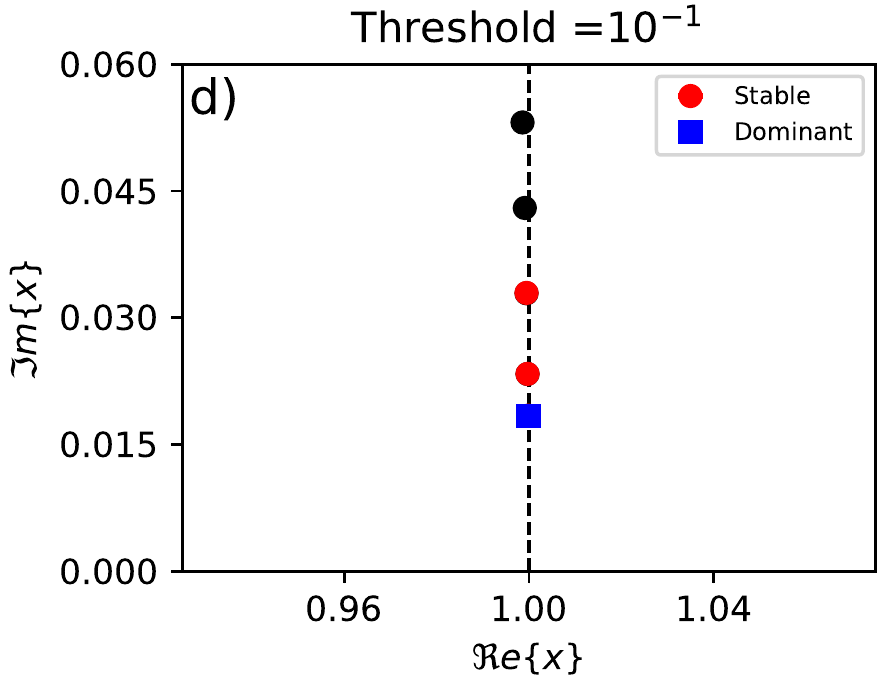}
 \caption{Map of zeros for different thresholds and the three most stable zeros (red circle), including the dominant one (blue square). The zero stability was calculated by generating $10$ new perturbed polynomials using equation \ref{pert} with $a=1\%$, finding the three most stable zeros and taking the mean value of its positions. The error bars are too small to be seen in the figure.}
 \label{all-threshold}
\end{figure}

	To solve this convergence issue some proprieties of the dominant zeros were explored. At this point it is worthy to note that we defined as dominant zero the zero closest to the point $(1,0)$. With that said, in a previous study \cite{Rodrigues2020}, it was observed that the dominant zero is stable to random perturbations in the polynomial coefficients, i.e., applying a perturbation in the EPD following the equation,

\begin{equation}\label{pert}
h^*(\beta_o) = h(\beta_o) \times (1 + \varepsilon a),
\end{equation}
where $h(\beta_o)$ is the EPD, $h^*(\beta_o)$ is the perturbed EPD, $\varepsilon \in [-1,1]$ is a uniform random variable and $a$ the perturbation, does not change the dominant zero position appreciably, as can be seen in figure \ref{est}. Even for different thresholds, the dominant zero is stable, although it is not the only stable zero, as can be seen in figure \ref{all-threshold}. Therefore, only the stability it is not enough to identify the dominant zero. In fact, the zeros which causes the convergence issue for the threshold of $10^{-2}$ are highly stable, as can be seen in figure \ref{all-threshold}. To correctly identify the dominant zero, it was added a bias to the algorithm to select among the most stable zeros, the zero closest to the line $x=1$ as the dominant one. Then, the following algorithm was proposed to avoid the selection of undesired zeros in automated application of the EPD zeros approach.

\begin{itemize}
    \item[1.] Apply a perturbation to the EPD following equation \ref{pert} and generate $m$ new polynomials.
    \item[2] Normalize $h_n(\beta_o)^*$, so its maximum value is $1$, and discard coefficients smaller than a given threshold.
    \item[2.] Find the zeros of those polynomials.
    \item[3.] Select the most stable zeros.
    \item[4.] Select the zero closest to the line $x = 1$ as the dominant zero.
\end{itemize}

\begin{figure}
 \centering 
 \includegraphics[width=0.49\columnwidth]{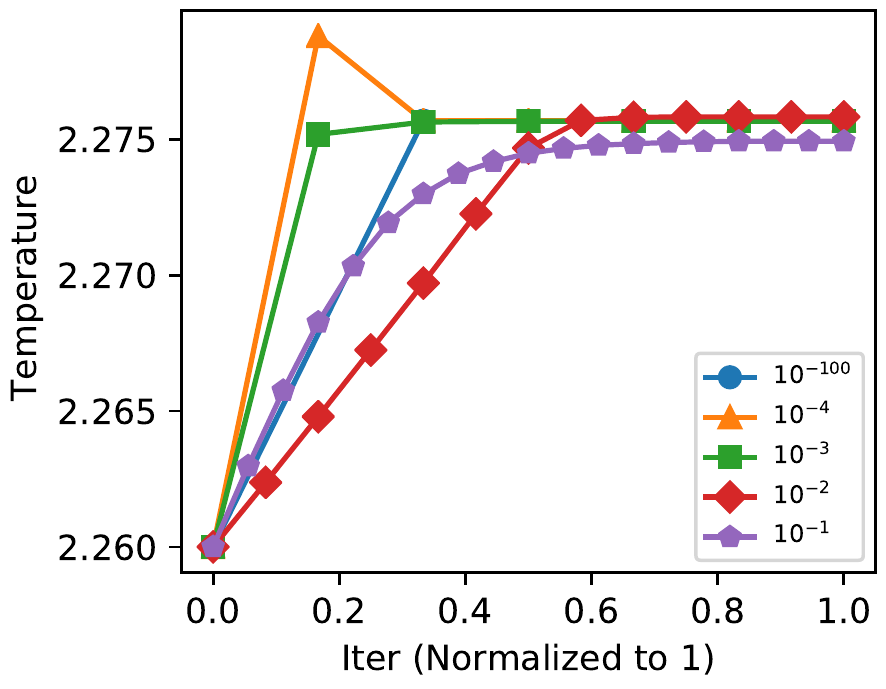}
 \caption{Process of convergence for all threshold, including the $10^{-2}$ after applying the new algorithm to select the dominant zero.}
 \label{all-conv}
\end{figure}

	With this, the dominant zero was consistently selected and the convergence was achieved in all simulations made; Figure \ref{all-conv} exemplifies our finding.

\subsection{\label{sec:potts} Six-states Potts model}

\begin{figure}[ht]
 \includegraphics[width=0.49\columnwidth]{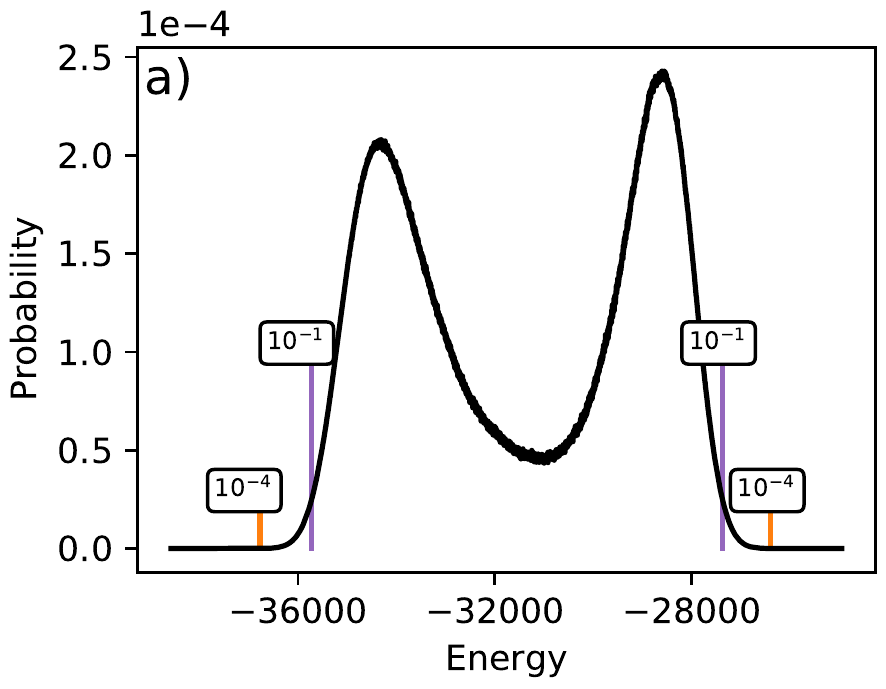}
 \includegraphics[width=0.49\columnwidth]{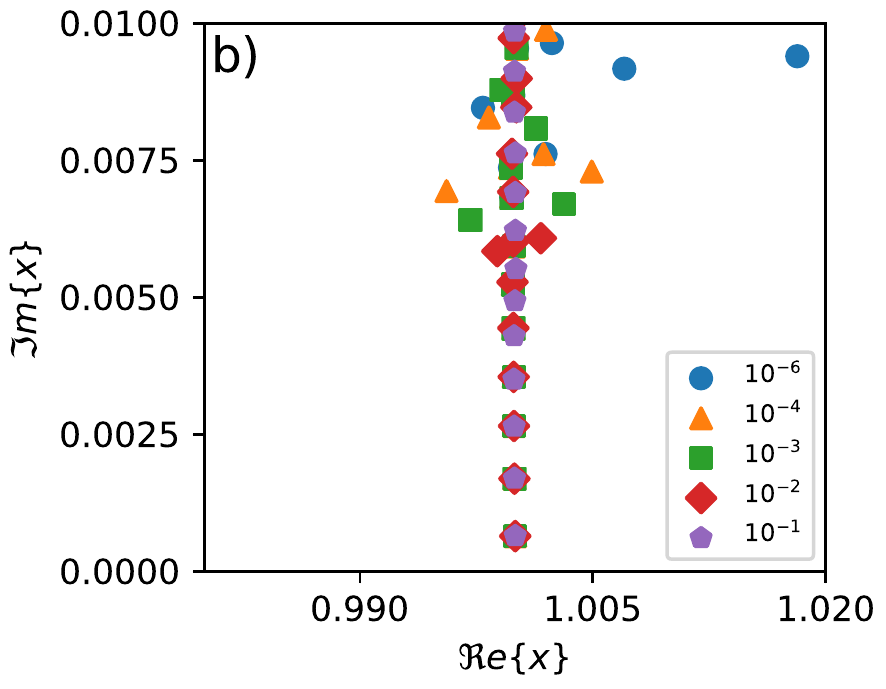}
 \caption{a) EPD for the six-states Potts model, made with $10^{8}$ MCS, with two lines indicating the position of the discard thresholds. Notice that, between the peaks are the vast majority of coefficients and only a few coefficients can be discarded using the chosen thresholds. b) Map of zeros for different thresholds where the dominant zero is easily identified.}
 \label{potts6-epd-zeros}
\end{figure}

The six-states Potts model has a more complex EPD due to its discontinuous phase transition, and more care is necessary at the discard step. As can be seen in figure \ref{potts6-epd-zeros}, the EPD has two large peaks, and as described in \cite{Costa2016a} states between those peaks that has low probability to occur should not be discarded. Therefore, even for higher threshold such as $10^{-1}$, since intermediate states have to be considered, almost no state is discarded. A reflection of such protection of states is seen in the critical temperature obtained for different threshold, table \ref{tab:tab-potts6-tc}, where no difference in $T_c$ is observed at any threshold. Despite this, all the values found for $T_c$ are close to the expected one of $T_c=0.807606J/k_b$ independent of the coefficients discard threshold and MCS that have been chosen.

In contrast to what we observed for the Ising model, no glimpse of convergence issues was observed in this system. Indeed, as is clear from figure \ref{potts6-epd-zeros}, there are no adjacent zeros to the dominant one in this case. Therefore, automated implementations are not expected to select undesired zeros.

\begin{table}[ht]
\centering
\begin{tabular}{l|ccc}
 \multicolumn{1}{c|}{Threshold}&\multicolumn{3}{c}{MCS} \\
  & $10^{6}$ & $10^{7}$ & $10^{8}$\\ \hline
 $10^{-6}$ & $0.80740(8)$ & $0.80736(4)$ & $0.80736(3)$ \\
 $10^{-4}$ & $0.80740(8)$ & $0.80736(4)$ & $0.80736(3)$\\
 $10^{-3}$ & $0.80740(8)$ & $0.80736(4)$ & $0.80736(3)$ \\
 $10^{-2}$ & $0.80740(8)$ & $0.80736(4)$ & $0.80736(3)$ \\
 $10^{-1}$ & $0.80740(8)$ & $0.80736(4)$ & $0.80736(3)$ \\
\end{tabular}
\caption{Critical temperature obtained for different combinations of threshold and Monte Carlo steps.}
\label{tab:tab-potts6-tc}
\end{table}

\begin{figure}[ht]
 \includegraphics[width=0.49\columnwidth]{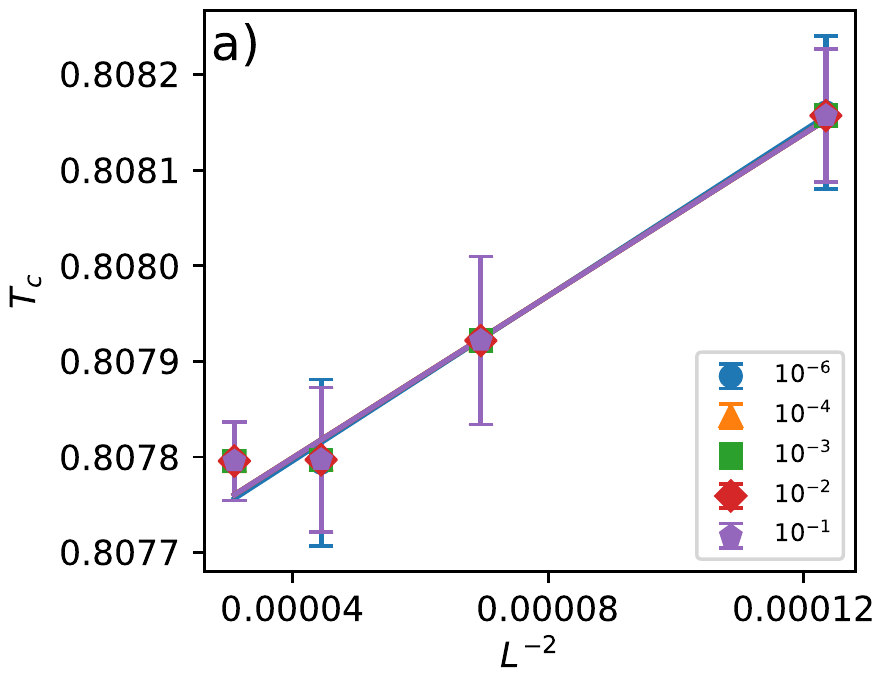}
 \includegraphics[width=0.49\columnwidth]{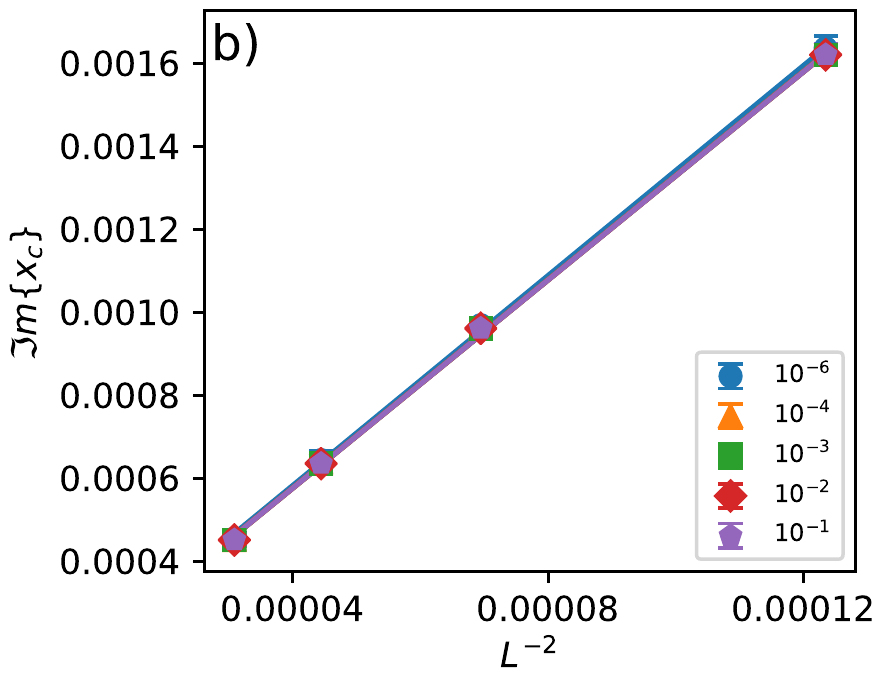}
 \caption{Linear regression to identify a) $T_c$ and the b) imaginary part of the dominant zero going to zero for discard thresholds ranging from $10^{-1}$ to $10^{-6}$. For clarity, the figure shows only results for $10^8$ MCS. The values found for $T_c$ were shown in table \ref{tab:tab-potts6-tc}.}
 \label{regression-potts6}
\end{figure}

One final remark about the six-state Potts model is that although in Ref.~\cite{Costa2016a} a simple linear adjustment did not work to find $T_c$, i.e., a correction to scaling had to be used to consistently estimate $T_c$, in this work, using larger lattice sizes, a correction to scaling it is not necessary. Indeed, while in Ref.~\cite{Costa2016a} the lattice sizes used were $L = [20, 40, 60, 80, 120]$ in this work we used $L = [90, 120, 150, 180]$. Thus, as can be seen in figure \ref{regression-potts6} the finite size scaling for the critical temperature and imaginary part of the dominant zero for larger sizes and the results obtained by the linear fits (see Table \ref{tab:tab-potts6-tc}) shows no need of corrections to scaling.

\section{\label{sec:conc}Conclusion}

	In this paper, we studied how the choice of the coefficients discard threshold and MCS used to estimate coefficients impact automated implementations of the EPD method and its capability to find the critical temperature and critical exponent. To do this, and thinking of expanding our results for continuous and discontinuous phase transitions, we used 2D Ising and six-state Potts models. The EPD method showed to be robust against the coefficients discard threshold and MCS used, given results for $T_c$ and $\nu$ close to the expected and that agree very well with each other even for considerable differences in the threshold and MCS. This allow considerable speedups in the estimation of transition temperatures since estimates are almost insensitive to reasonable choice of the parameters. Indeed, results obtained using 10$^7$ or 10$^8$ MCS agree withing error bars with minimal accuracy difference. This means that about 10 times less computational effort is necessary to estimate critical temperature and exponent with almost the same accuracy. As expected, we also find that the correction to scaling used in Ref. \cite{Costa2016a} for the first order transition of the six-states Potts model is not necessary for larger lattice sizes. 
	
	Despite the robustness of the EPD method, the Ising model showed a convergence issue for a specific choice of the coefficients discard threshold, $10^{-2}$, due to a misleading automated choice of the dominant zero. This behavior led to an oscillating pattern in the iteration of the EPD algorithm instead of a convergent one. To solve this issue, a bias was added to the algorithm to correctly select the dominant zero and ignore other adjacent zeros nearby that may be mistakenly chosen. Although this corrects the issue, we must point out that this oscillatory behavior for the automated convergence of the Ising model was found for this specific choice of coefficients discard threshold. The solution present here is not unique and it may even be much more advantageous to reduce the number of zeros by using higher coefficients discard thresholds without significant impact in the final estimates of the critical temperature and exponent than employing the algorithm to deal with this oscillatory behavior. Indeed, it is remarkable how the estimates are almost insensitive to variations on the coefficients discard threshold and on the number of MCS considered in the building of the EPD and map of zeros. Of course, for more complicated models, the algorithm presented here may be very useful in preventing this oscillatory behavior. 

\section*{Acknowledgements}
The authors gratefully acknowledge the financial support from CNPq grant $402091/2012-4$ and FAPEMIG grant RED$-00458-16$.


\bibliography{mybibfile}

\end{document}